\documentstyle[emulateapj,10pt,apjfonts,lineno,psfig]{article}

\def\psr{B1823--13}
\def\xmm{{\em XMM-Newton}}

\newcommand\kms{km~s$^{-1}$}
\newcommand\etal{{\rm et~al.\ }}
\renewcommand\farcm{\hbox{$.\!\!^{\prime}$}}

\lefthead{GAENSLER ET AL.}
\righthead{\uppercase{XMM Observations of PSR~\psr}}

\begin{document}
\title{XMM-Newton Observations of PSR~B1823--13: An Asymmetric Synchrotron Nebula
Around a Vela-like Pulsar}
\author{B. M. Gaensler\altaffilmark{1}, N. S. Schulz\altaffilmark{2},
V. M. Kaspi\altaffilmark{3,2}, 
M.~J.~Pivovaroff\altaffilmark{4} and
W. E. Becker\altaffilmark{5}}
\altaffiltext{1}{Harvard-Smithsonian
Center for Astrophysics, 60 Garden Street MS-6, Cambridge, MA 02138;
bgaensler@cfa.harvard.edu}
\altaffiltext{2}{Center for Space Research, Massachusetts Institute
of Technology, 70 Vassar Street, Cambridge, MA 02139}
\altaffiltext{3}{Physics Department, McGill University,
3600 University Street, Montreal, Quebec, Canada}
\altaffiltext{4}{Space Sciences Laboratory,
University of California, Berkeley, CA 94720}
\altaffiltext{5}{Max-Planck-Institut f\"{u}r Extraterrestrische Physik, D-85740 Garching, Germany}
\hspace{-5cm}

\begin{abstract}

We present a deep observation with the {\em X-ray Multi-Mirror
Mission}\ of PSR~\psr, a young pulsar with similar properties
to the Vela pulsar.
We detect two components
to the X-ray emission associated with PSR~\psr: an elongated
core of extent $30''$ immediately surrounding the pulsar, embedded in a fainter,
diffuse component of emission $5'$ in extent, seen only on the
southern side of the pulsar. The pulsar itself is not detected, either
as a point source or through its pulsations. Both components of the
X-ray emission are well fit by a power law spectrum, with photon index
$\Gamma \approx 1.6$
and X-ray luminosity (0.5--10~keV) $L_X
\approx 9\times10^{32}$~erg~s$^{-1}$ for the core, and
$\Gamma \approx 2.3$ and $L_X \approx 3\times10^{33}$~erg~s$^{-1}$ for
the diffuse emission, for a distance of 4~kpc.
We interpret both components of emission as corresponding to a
pulsar wind nebula, which we designate G18.0--0.7.
We argue that the core region 
represents the wind termination shock of this nebula,
while the diffuse component indicates the shocked downstream wind. We
propose that the asymmetric morphology of the diffuse emission with
respect to the pulsar is the result of a reverse shock from an
associated supernova remnant, which has compressed and distorted the
pulsar-powered nebula. Such an interaction might be typical
for pulsars at this stage in their evolution.
The associated supernova remnant is not detected
directly, most likely being too faint to be seen in
existing X-ray and radio observations.

\end{abstract}


\section{Introduction}
\label{sec_intro}

Many of the energetic processes which take place in the aftermath of a
supernova explosion are significant sources of X-ray emission. The
supernova remnant (SNR) which results as the expanding debris collide
with the interstellar medium (ISM) can generate both thermal X-rays
from shock-heated gas and ejecta, as well as non-thermal X-ray
synchrotron emission from the shock-acceleration of electrons.
Meanwhile, the core collapse which triggers such explosions often forms
a rapidly rotating central pulsar.  Pulsars also produce X-rays, in
the form of pulsed non-thermal emission from the pulsar magnetosphere,
quasi-blackbody emission from the neutron star surface, or an extended
pulsar wind nebula (PWN) powered by a relativistic particle outflow
from the central source.

Each of these processes dominates at different stages in the system's
evolution. For example, the youngest systems are generally well-suited for
studying SNR blast waves and magnetospheric emission from pulsars,
while in middle-aged systems one can better study the neutron star surface.
Potentially the most challenging group of objects to understand are the
``Vela-like'' pulsars (\cite{bt97}), which are at an evolutionary stage
for which {\em all}\ the above processes may be taking place. Such
pulsars, typified by the Vela pulsar, have spin periods $P\sim100$~ms,
spin-down luminosities $\dot{E} \equiv 4\pi^2I\dot{P}/P^3\sim
10^{36}-10^{37}$~erg~s$^{-1}$ and characteristic ages $\tau \equiv P/2\dot{P}
\sim 10-20$~kyr (where $\dot{P}$ is the pulsar's
spin-period derivative and $I$
is the star's moment of inertia).  Until recently,
about 10 Vela-like pulsars were known; the highly successful
Parkes Multibeam Pulsar Survey has approximately doubled this number
(\cite{mbc+02}).

The Vela pulsar itself is sufficiently nearby ($\sim250$~pc) that its
different emission processes can be clearly distinguished: X-rays from
the neutron star surface, the pulsar magnetosphere, the surrounding PWN
and the associated SNR have all now been identified (\cite{shd99};
\cite{la00}; \cite{hgh01}; \cite{pzs+01}). However, the rest of the
known Vela-like pulsars are at much greater distances,
resulting in fainter, smaller and more heavily absorbed
X-ray sources. It has thus been difficult to detect and interpret the
emission from these objects.

Radio observations of the pulsar \psr\ have measured a spin-period
$P=101$~ms and a period derivative $\dot{P} =
7.5\times10^{-14}$~s~s$^{-1}$, implying a spin-down luminosity $\dot{E}
= 2.8\times10^{36}$~erg~s$^{-1}$ and a characteristic age $\tau =
21.4$~kyr (\cite{clj+92}), all properties typical of a Vela-like
pulsar.  The distance to PSR~\psr, as inferred from the dispersion of
its radio pulses, is $3.9\pm0.4$~kpc (\cite{cl02}).  Previous X-ray
studies of PSR~\psr\ have been carried out with both {\em ROSAT}\ and
{\em ASCA}. The {\em ROSAT}\ data showed a complicated source at the
pulsar's position, with a morphology best modeled as a point source
embedded in a compact nebula of radius 20$''$ and surrounded by an
elongated diffuse region $\sim$5$'$ across (\cite{fsp96}).  However,
because of the low number of photons detected, it was not possible to
determine whether the extended component of the source was a SNR or
PWN, while the nature of emission from the pulsar itself also remained
unclear. {\em ASCA}\ data also indicated the presence of one or more
compact components embedded in a more extended region of emission, but
lacked the spatial resolution needed to properly separate and interpret
these sources (\cite{skts01}).

As the first part of an effort to understand the X-ray
emission from Vela-like pulsars, here we present observations of pulsar
\psr\ with the {\em X-ray Multi-Mirror Mission}\ (\xmm), the first observatory
to provide sufficient angular, spectral and temporal resolution to
separately identify all the processes described above. \xmm\ also has
much higher sensitivity (effective area 4650~cm$^2$ at 
1.5~keV) than previous missions, making
it well-suited for studying faint sources such as seen here.

\section{Observations and Analysis}
\label{sec_obs}

Observations of PSR~\psr\ were carried out with \xmm\ on 2001 October 16
and 2001 October 18, in two observations each of length $\sim$30~ks,
as summarized in Table~\ref{tab_obs}.
The data described here correspond to the three X-ray imaging
instruments on board \xmm. 
The EPIC MOS1 and MOS2
detectors were operated in the standard ``full frame'' mode, in which
seven CCDs in each detector are used to produce an approximately
circular field-of-view of diameter $30'$, with a time resolution for
each CCD frame of 2.6~s.  The EPIC pn detector was operated in ``small
window'' mode, in which only a central $4'\times4'$ square region is
active, but for which the time resolution is 5.7~ms. In this mode a
significant fraction of each 5.7-ms frame is used to read out the CCD,
resulting in a dead-time fraction of 30\% (\cite{kbkb99}).  To avoid optical
contamination of the field, the medium and thin blocking filters were
used for the MOS and pn detectors, respectively.

Initial processing of the data was carried out at the \xmm\ Science
Operations Centre (SOC). We analyzed the resulting event files using
the \xmm\ Software Analysis System (SAS), version 5.3.0.  The data were
first filtered to remove hot pixels and other bad data,
and to only allow standard event grades (patterns 0 to
12 for MOS1 and MOS2, patterns 0 to 4 for pn).  Each of these data sets
was examined for periods of high background by only considering events
with energies in the range 10--15~keV.  Times at which flares or high
levels were seen in the X-ray count rate for this bandpass were
excluded.  This latter filtering excluded a significant fraction of
each observation; the resulting useful exposure times are listed in
Table~\ref{tab_obs}.

\subsection{Imaging}
\label{sec_obs_imag}

For each observation and detector listed in Table~\ref{tab_obs}, the
data set was corrected was for vignetting losses using the SAS task
EVIGWEIGHT. Next, an energy filter was applied to include only events
falling in the energy range 0.5--10~keV, other energies being dominated
by background. Finally, all the MOS and pn data were combined to form a
single image for each type of detector.  For the MOS CCDs, the
flat-fielded images have a plate scale of $1\farcs1$~pixel$^{-1}$;  for
the pn CCD, the flat-fielded image has a plate scale of
$4\farcs4$~pixel$^{-1}$.

\subsection{Spectroscopy}
\label{sec_obs_spec}

The high background count-rate for \xmm\ requires careful analysis in
order to extract accurate source spectra.  The background contribution
to the data consists of two main components (see \cite{lwpd02}): a
diffuse X-ray background, assumed to have a smooth spatial distribution
over the observed field, and a particle background, which shows spatial
variations across the detector.

For a source not much larger than the point spread function (PSF), we
determine the spectrum by  extracting data from  both a small region
enclosing the source, and from a reference spectrum immediately
adjacent. The spectrum of interest is then obtained by subtracting the
reference spectrum from the source spectrum, scaling appropriately to
account for the differing areas of the extraction regions for the two
fields.  This approach assumes that because of the proximity of the two
extraction regions, the differences in the background components are
negligible between source and reference spectra.

Analysis of extended sources is complicated by the spatially varying
nature of the background.  Both the diffuse X-ray background and the
source photons are vignetted by the \xmm\ mirrors, resulting
in a background component that varies dramatically with detector
position.   The particle background, however, is unaffected by the
mirrors, resulting in a background component that is largely
independent of detector position.  Given these complications, we follow
the prescription given in Appendix~A of Arnaud
\etal\ (2002\nocite{aml+02}), in which reference spectra from adjacent
regions are used to correct for the X-ray background, and spectra from
blank-field observations supplied by the
SOC\footnote{http://xmm.vilspa.esa.es/ccf/epic/\#background} are used
to account for the particle background.

Once these corrections had been applied, spectra were re-grouped so
that there were at least 50 counts per spectral bin for compact
sources, and 100 counts per bin for extended sources.  For analysis of
EPIC MOS data, we used the responses supplied by the
SOC\footnote{http://xmm.vilspa.esa.es/ccf/epic/\#responses} to provide
information on the redistribution matrix and effective area of each
CCD: specifically, for the MOS1 CCD we used the response file {\tt
m1\_medv9q19t5r5\_all\_15.rsp}, while for MOS2 we used {\tt
m2\_medv9q19t5r5\_all\_15.rsp}. For analysis of EPIC pn data, we
generated our own response files using the SAS tasks RMFGEN and ARFGEN.
Subsequent spectral fitting and analysis were carried out using XSPEC
version 11.1.0.

\subsection{Timing}

In the ``small window'' mode used here,
the EPIC pn CCD has sufficient time resolution to search for X-ray
pulsations from PSR~\psr. Due to improvements in the time-tagging of events
since the time when pipeline processing was carried out on the data by
the SOC, we completely reprocessed the pn data from the raw observation
data files, using the SAS task EPCHAIN. After  filtering the data as
described in \S\ref{sec_obs} above, we then corrected the arrival
times of each event to correspond to a reference frame at the solar system
barycenter. For a given extraction region and energy range, a pulsation
analysis could then be carried out on the resulting arrival times.

\section{Results}
\label{sec_results}

\subsection{Imaging}

\begin{figure}[hbt]
\centerline{\psfig{file=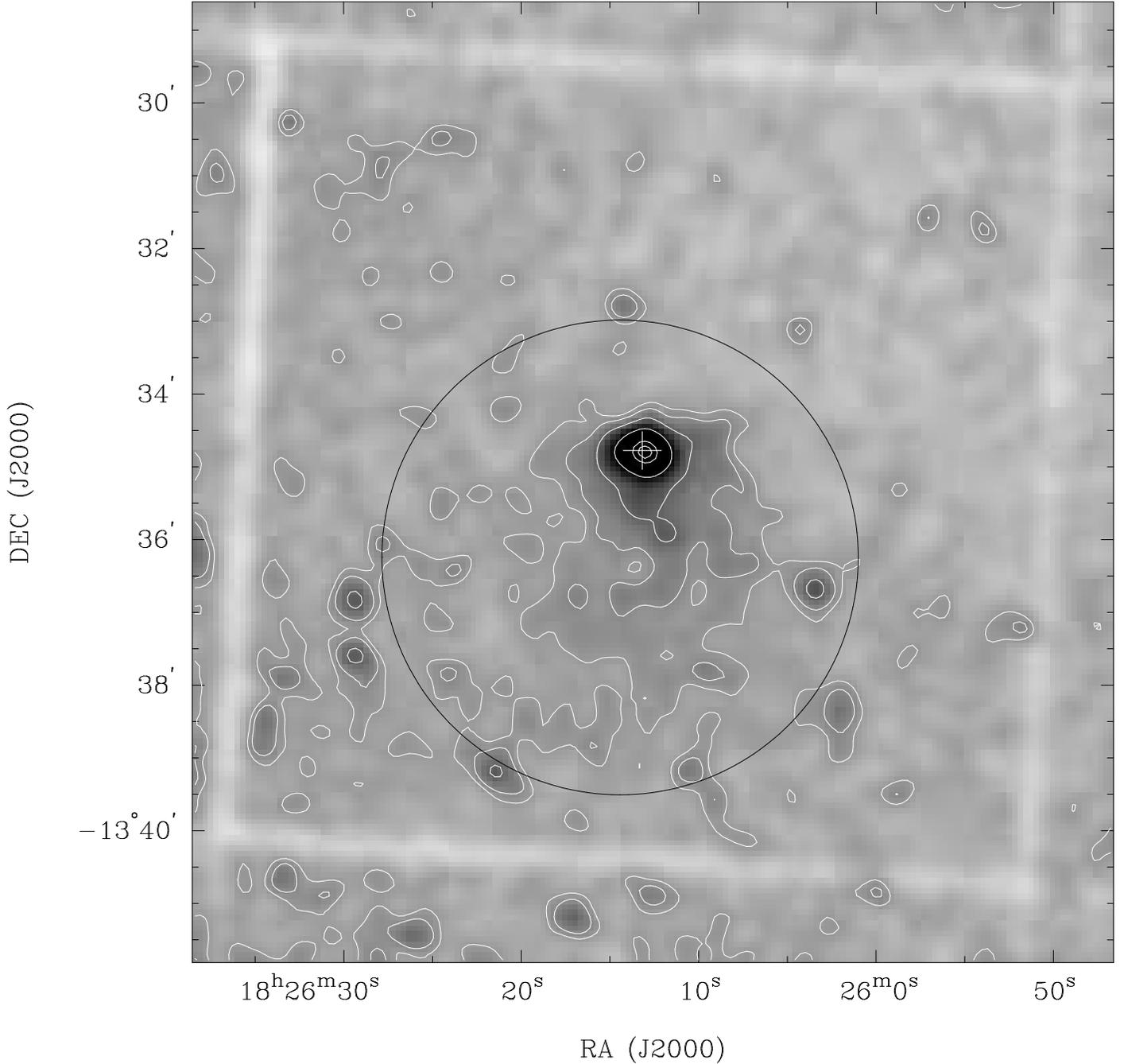,width=\textwidth,angle=270}}
\caption{EPIC MOS image of the field surrounding PSR~\psr, produced by
combining all MOS1 and MOS2 data listed in Table~\ref{tab_obs} in the energy range
0.5--10~keV. The image has been corrected
for off-axis vignetting, and then convolved with a gaussian of FWHM $20''$.
The image is displayed
using a linear transfer function ranging between 0\%
and 30\% of the peak intensity in the field; the white
contours are plotted at levels of 13\%, 15\%, 20\%, 35\%,
75\% and 90\% of the peak.  
The white cross marks the position of the pulsar as
listed in Table~\ref{tab_psr}, while the
circle shows the outer edge of the extraction region used
for the spectrum of the diffuse component of the source.
Linear structures in the image correspond to gaps
between individual CCDs.}
\label{fig_mos_all}
\end{figure}

\begin{figure}[hbt]
\centerline{\psfig{file=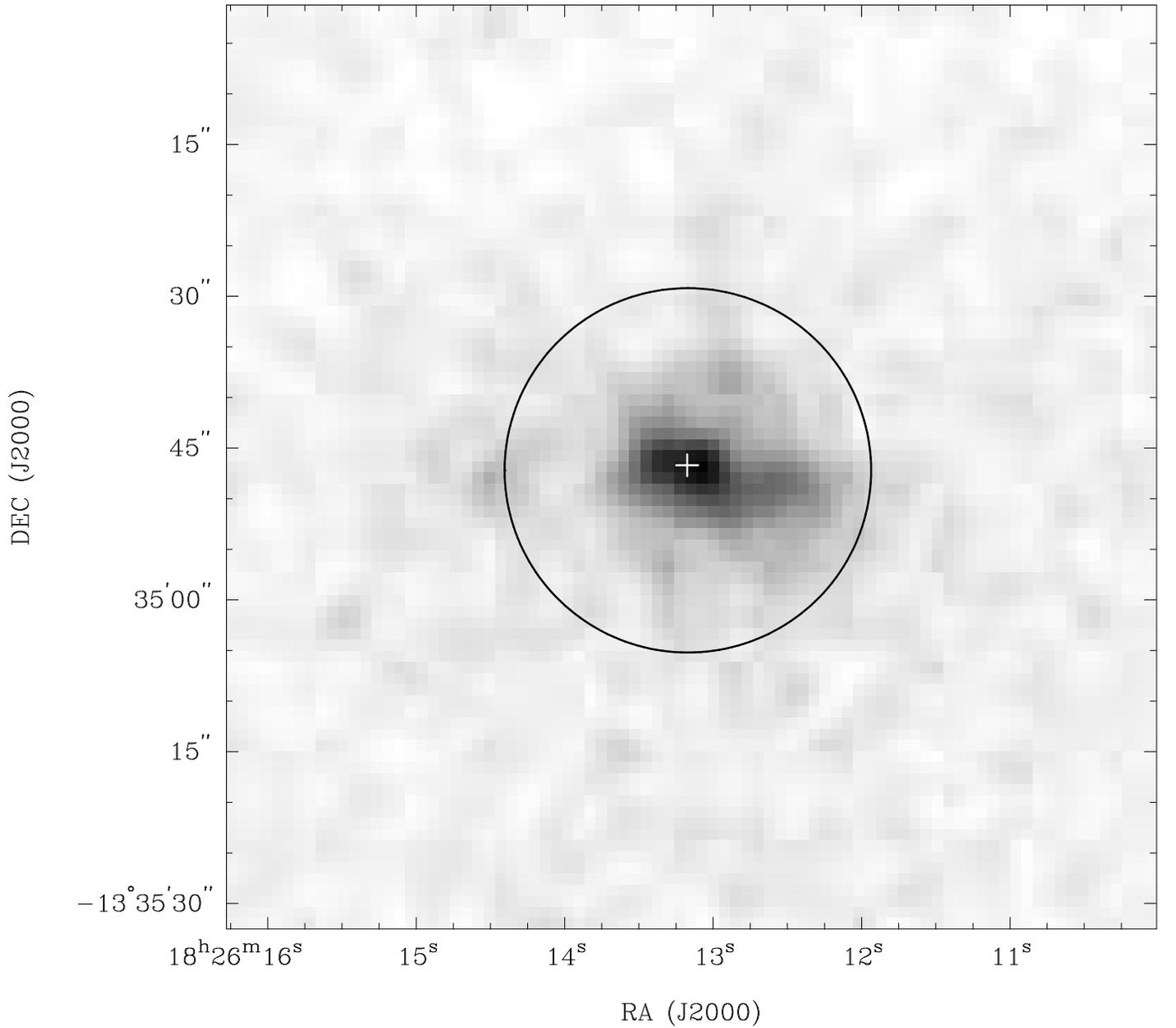,width=\textwidth,angle=270}}
\caption{As in Figure~\ref{fig_mos_all},
but showing emission from the central
$90\arcsec\times90\arcsec$ region surrounding
the pulsar, and smoothed with a gaussian
of FWHM $3''$.  The image shows a linear transfer function
ranging between 0\% and 100\% of the peak intensity.
The white cross  marks the position of the pulsar as
listed in Table~\ref{tab_psr}, while the circle shows the extraction
region for the spectrum of the core.}
\label{fig_mos_zoom}
\end{figure}

The EPIC MOS image of PSR~\psr\ is shown in Figures~\ref{fig_mos_all}
and \ref{fig_mos_zoom}.
Because a significant contribution to the
background emission is from high-energy particles, the vignetting
correction which has been applied to the data gives the false
impression that the background emission is stronger around the edges of
the fields. The O~star HD~169727 (outside the extent of the region
shown in Figure~\ref{fig_mos_all}) is detected as a prominent X-ray
source; comparison of its position in our data with that in the Tycho
Reference Catalogue (\cite{hkb+98}) suggests that the absolute astrometry of our
observation is correct to within $0\farcs5$ in each coordinate.

\begin{figure}[hbt]
\centerline{\psfig{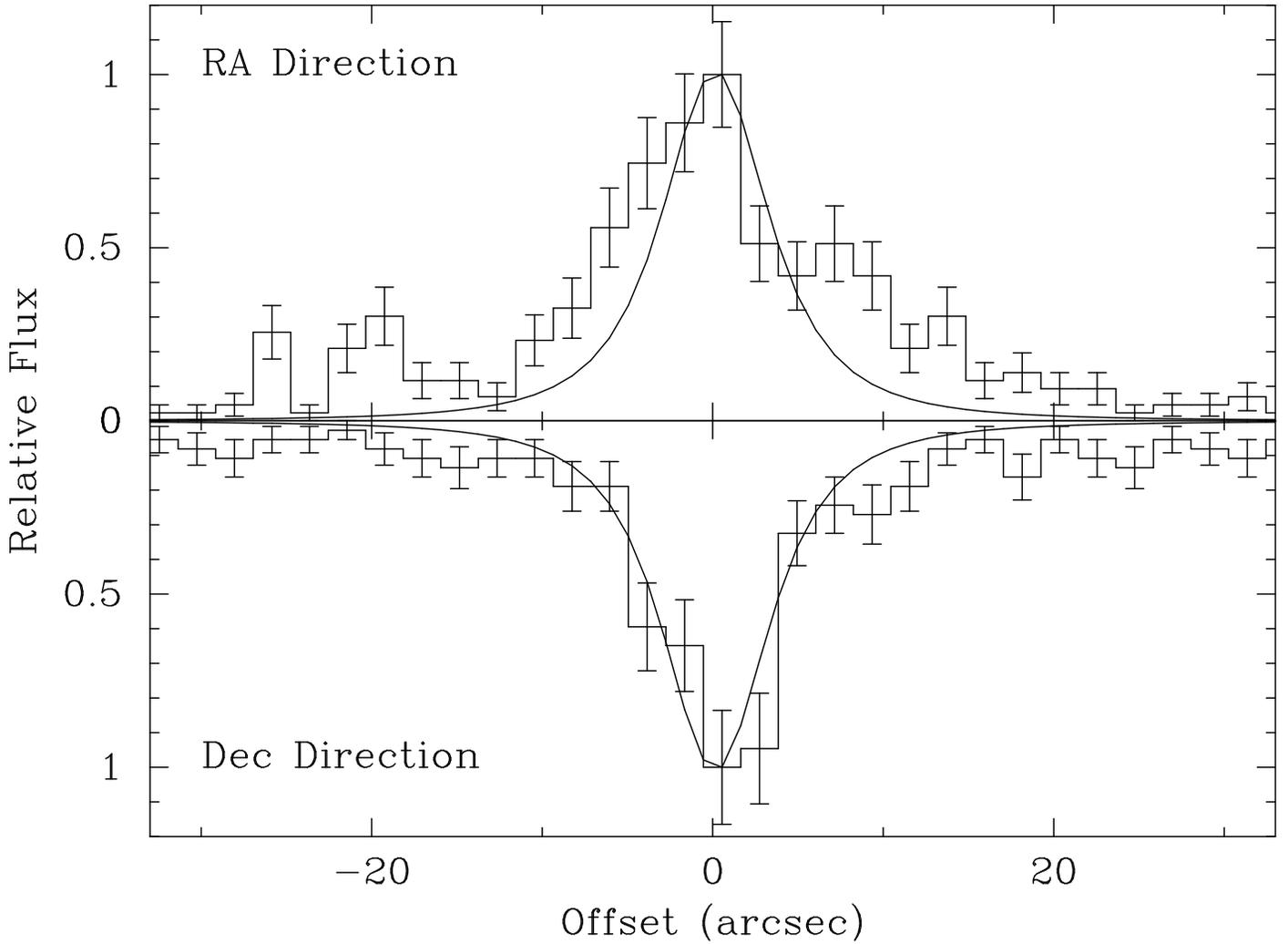}}
\caption{Intensity profiles 
through the core region of G18.0--0.7. The histograms
plot the summed flux of EPIC MOS data in
successive binned regions along each of the RA and Dec axes.
The size of each region is two MOS pixels ($2\farcs2$) along the axis
of offset, and three MOS pixels ($3\farcs3$) along the axis for which
the data are summed. Positive offsets along the RA axis (upper panel)
correspond to more westerly points, while positive offsets along
the Declination axis (lower panel) correspond to more northerly data;
in both cases the pulsar position is at zero offset.
The smooth curve on each plot is the profile
of the EPIC MOS point spread function at an energy of 1.5~keV, after
similar summing as applied to the image.}
\label{fig_core_slice}
\end{figure}

The data clearly demonstrate that the X-ray emission associated with
PSR~\psr\ consists of at least two components.  The brightest X-ray
emission is seen immediately surrounding the position of the pulsar, in
a compact region shown in Figure~\ref{fig_mos_zoom} which we
subsequently refer to as the ``core''.  However, the morphology of the
core suggests that it is not an unresolved source, but rather is
elongated along an extent of $\sim30''$ in the east-west direction.
This extension can be seen independently but at lower resolution in the
EPIC pn image (not shown here).  We quantify this in
Figure~\ref{fig_core_slice}, where we show the intensity profile of
X-ray emission around the pulsar along both east-west and north-south
axes. We compare these profiles to the \xmm\ PSF for MOS1 at 1.5~keV, 
which we generated using the King profile
parameters included
in the \xmm\ calibration file {\tt XRT1\_XPSF\_0005.CCF}.\footnote{See
http://xmm.vilspa.esa.es/docs/documents/CAL-SRN-0100-0-0.ps.gz 
for more information.}
Above levels of $\sim25\%$ of the peak flux in the image, the
intensity profile in the north-south direction (i.e.\ the lower panel
of Fig~\ref{fig_core_slice}) appears to be largely consistent with this
PSF.  However, the profile shown in the upper panel
of Figure~\ref{fig_core_slice} clearly demonstrates that along its
east-west axis, the core region is significantly broader than the PSF,
only falling to the apparent background level at offsets of
$\approx15''$ on the eastern and western sides of the pulsar.
Also present is a secondary clump of emission $20''$ east
of the pulsar, seen at coordinates (J2000) RA $18^{\rm h}26^{\rm m}14\fs5$,
Dec.  $-13^\circ34'47''$ in Figure~\ref{fig_mos_zoom}, and 
and visible as a flux enhancement at an offset
of $-20''$ in the upper panel of Figure~\ref{fig_core_slice}.  

\begin{figure}[hbt]
\centerline{\psfig{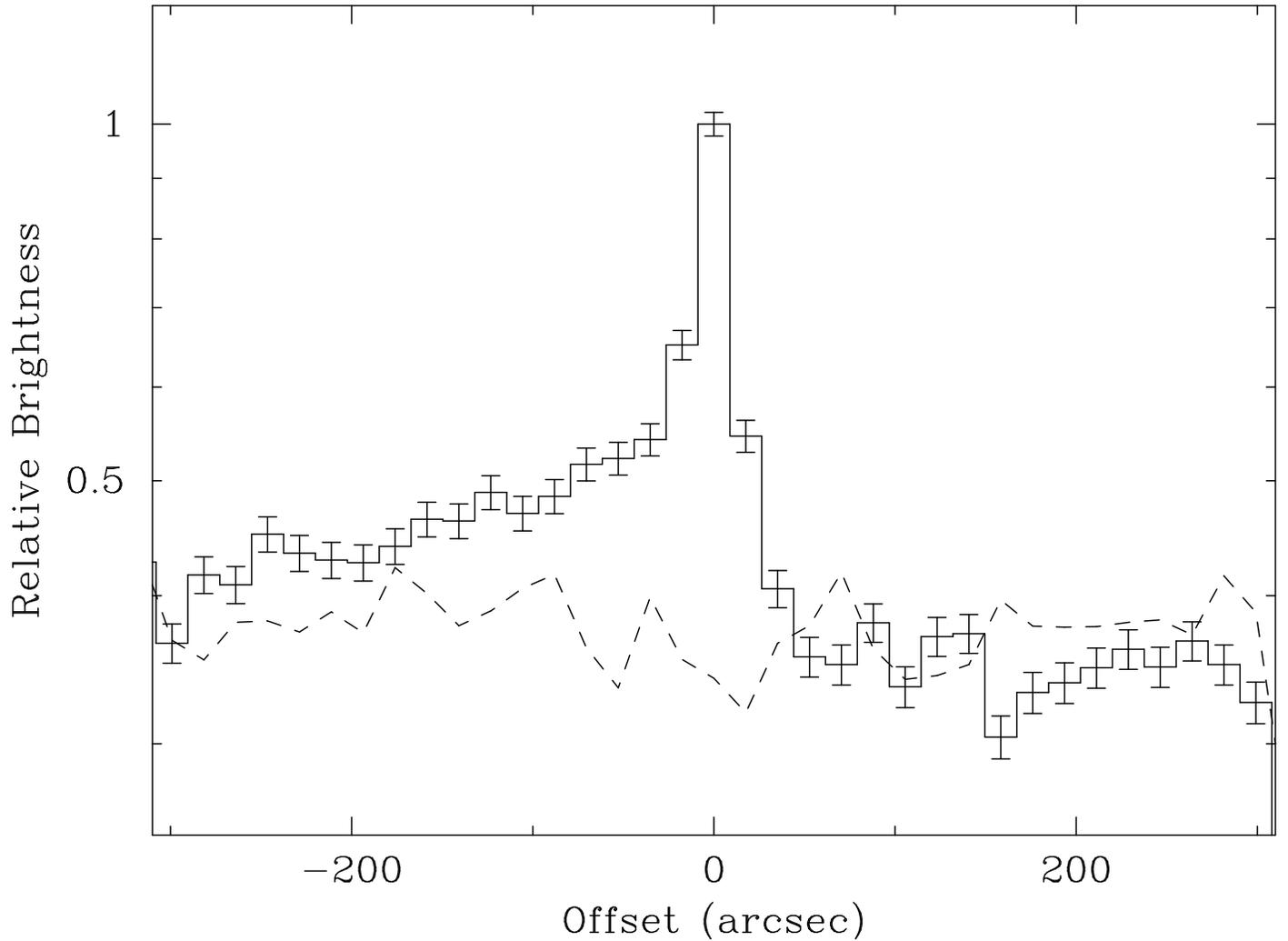}}
\caption{Intensity profiles through G18.0--0.7 along 
the north-south direction. The solid histogram
plots the summed flux of EPIC MOS data in successive binned
regions along the Declination axis. The size of each region is 
16 pixels ($17\farcs6$) along the offset axis,
and 208 pixels ($3\farcm8$) along the axis for which
the data are summed. Increasing offsets correspond to
more northerly data; the pulsar position is at zero offset.
The broken line corresponds to an identical analysis for
blank field observations, after
appropriate normalization (see text for details).}
\label{fig_pwn_profile}
\end{figure}

At larger angular scales, Figure~\ref{fig_mos_all} demonstrates the
existence of a faint, amorphous component of emission extending to much
larger distances from the pulsar than does the central core; we
subsequently refer to this region as the ``diffuse'' component.  This
diffuse emission is located predominantly to the south of the core
region. The morphology of the diffuse component is approximately
elliptical, centered $\sim1\farcm5$ south of
the pulsar, with a major axis of $\sim5'$ aligned north-south,
and a minor axis of $\sim4'$.
In all directions, the diffuse component decreases in
brightness with increasing distance from the pulsar. To the south, east
and west of the pulsar, the diffuse component fades into the background
over several arcmin; to the north of the pulsar the brightness profile
decreases much more rapidly. We demonstrate this in
Figure~\ref{fig_pwn_profile},  where we show the brightness profile of
the X-ray emission along the north-south axis of the source. As a
comparison, we have extracted the same profile from the blank-field
observations supplied by the SOC, and have normalized the profile
for the blank-field data to that of the source by comparing count rates
in an identical source-free region. As shown in
Figure~\ref{fig_pwn_profile}, a comparison of profiles between the
source field and the blank field clearly shows that X-ray emission from
the source extends at least $300''$ to the south of PSR~\psr, but
extends no more than $\sim40''$ to the north of the pulsar. Note that
the ranges of offset shown in Figure~\ref{fig_pwn_profile} 
correspond to the extent of
the central CCD of the detector. At larger offsets the background
dominates, and in these regions the differences between the background
levels in the source field and in the blank field make it impossible to
determine whether there is any significant signal above the noise.
For this reason we cannot  place useful constraints on the presence of
a faint limb-brightened shell of emission at larger radii, as might
be produced by an associated SNR (see further discussion
in \S\ref{sec_no_snr} below).

\subsection{Spectroscopy}

We have extracted spectra for the core from both the MOS and pn data,
using a circular extraction region, centered on the pulsar and with a
radius of $18''$ (as shown in Figure~\ref{fig_mos_zoom}).  Although
only $\sim$75\% of the energy from an unresolved or slightly extended
source is encircled at this radius, this extraction region minimizes
contamination from the surrounding diffuse component.  The core region
is sufficiently compact that we can correct for background using the
simple process of subtracting a reference region from an adjacent
source-free region. In this case, the reference region was located
$\sim3'$ distant from the core, and was placed so as to avoid any
contribution from the diffuse component of the emission.

Six spectra
resulted, one for each of MOS1, MOS2 and pn, for each of the two
observations listed in Table~\ref{tab_obs}.  We then fitted each of power-law
and blackbody models to these six spectra simultaneously, modified
by photoelectric absorption ---
the best-fit results are summarized in Table~\ref{tab_spec}. 
We find that the core region
is well fit by a power law, with an absorbing column of
$N_H = (1.0\pm0.2)\times10^{22}$~cm$^{-2}$ and a photon index $\Gamma =
1.6^{+0.1}_{-0.2}$. At a distance to the system of
$4d_4$~kpc, the unabsorbed 0.5--10~keV luminosity
implied by this fit is $L_X \approx 9d_4^2\times10^{32}$~erg~s$^{-1}$,
where we have corrected for the $\sim25\%$ of the flux from this
source which presumably falls outside our extraction region.
A blackbody model gives a much poorer fit to the spectrum.

\begin{figure}[hbt]
\centerline{\psfig{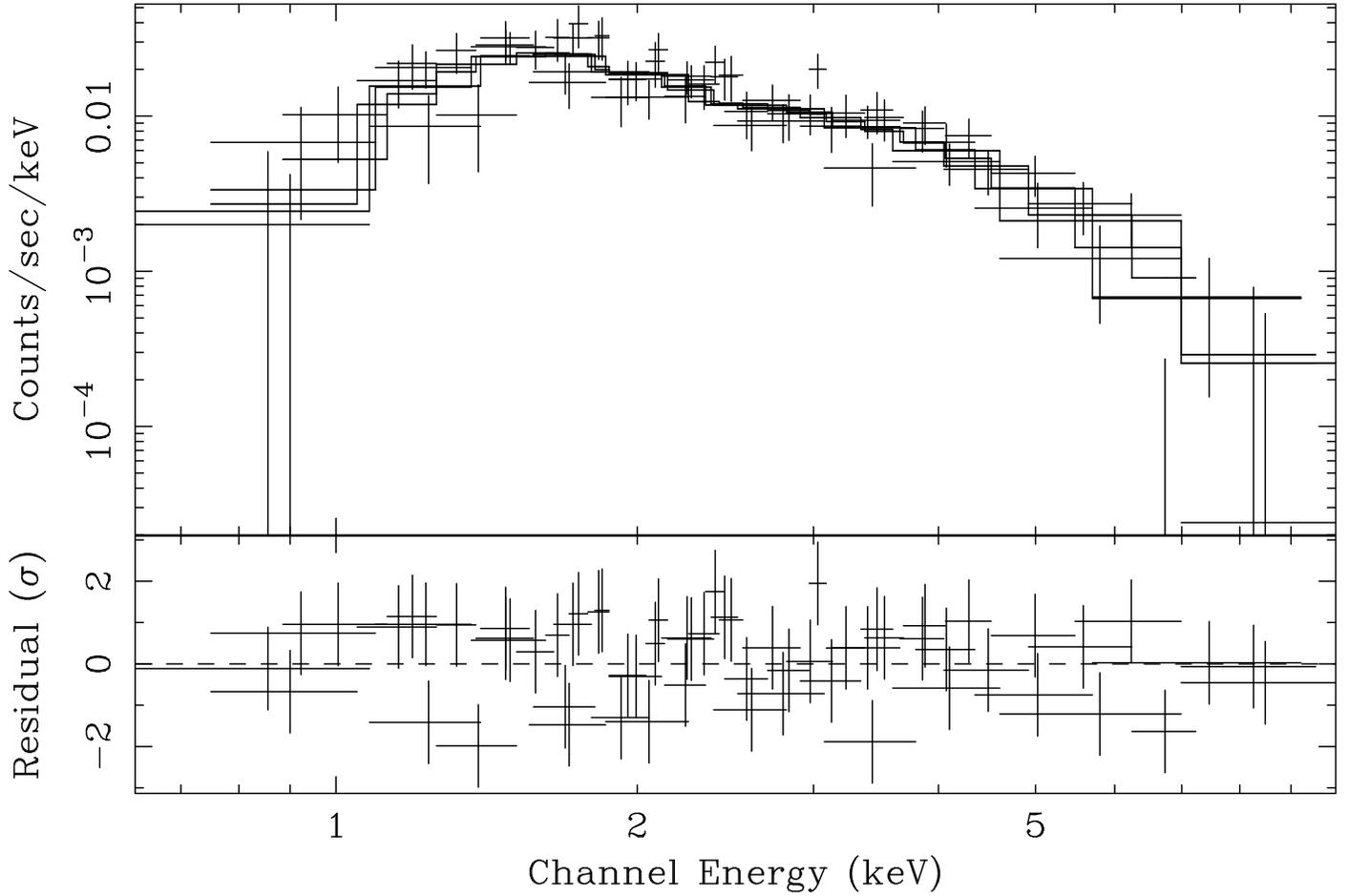}}
\caption{ \xmm\ spectrum of the diffuse component of G18.0--0.7.  The
data points in the upper panel correspond to the data from each of the
MOS1, MOS2 detectors for each of the two observations, while the solid
lines show the corresponding best-fit absorbed power-law model. The
lower panel shows the number of standard deviations by which the model
and the data differ in each bin.  The data have been plotted so as to
give a signal-to-noise ratio of at least 3 in each bin.}
\label{fig_pwn_spec}
\end{figure}

The diffuse component occupies a much larger area on the detectors than
does the core region, and indeed fills the entire field-of-view of the
EPIC pn image.
Because there is no source-free region on the
detector, it is impossible to accurately account for the background
of the diffuse emission for the EPIC pn. Instead, we only
considered MOS data in this analysis.
We extracted a spectrum from a circular region centered on
(J2000) RA $18^{\rm h}26^{\rm m}14\fs4$, Dec.\ $-13^\circ36'14\farcs8$, 
with radius $200''$, as shown
in Figure~\ref{fig_mos_all}.  We excluded from this region a
$30''$-radius region centered on the pulsar, and $20''$-radius regions
around the point sources seen at (J2000) RA $18^{\rm h}26^{\rm m}03\fs5$, Dec.\ 
$-13^\circ36'41\farcs7$
and RA $18^{\rm h}26^{\rm m}10\fs6$, Dec.\ $-13^\circ39'08\farcs4$. 
As a reference spectrum, we
considered all emission outside the main extraction region, but falling
within a square of side $9\farcm5$ centered on the pulsar.
Both source and reference spectra were corrected for the particle
background by subtraction of blank field data, as explained in 
\S\ref{sec_obs_spec}.
All four MOS spectra were then fit simultaneously to a variety
of absorbed models (see Table~\ref{tab_spec}
for details): a power law, thermal bremsstrahlung, 
a synchrotron spectrum with an exponential cut-off
(model ``SRCUT''), and three thermal plasma models
(``RAYMOND'', ``MEKAL'' and ``NEI'').
In all six cases,
we find a good match to the data with an absorbing column
of $N_H \sim 1.2 \times 10^{22}$~cm$^{-2}$. For the
power law fit shown in Figure~\ref{fig_pwn_spec}, 
the photon index is $\Gamma \approx 2.3$, while
the thermal models all imply temperatures $kT \sim 3.5$~keV. 
The cut-off synchrotron spectrum can be extrapolated to
radio wavelengths, where the corresponding 1~GHz
flux density implied for the source is $15\pm1$~mJy.\footnote{1 mJy
$= 10^{-29}$~W~m$^{-2}$~Hz$^{-1}$.} 

\subsection{Limits on a Point Source}
\label{pt}

While it is clear that the core component of X-ray emission surrounding
PSR~\psr\ is extended, an important consideration is how much of the
X-ray emission detected could be from an unresolved source potentially
corresponding to the pulsar itself. We here consider the upper limits
we can place on the temporal, morphological and spectral contributions
of a central point source.

\subsubsection{Timing Analysis}
\label{pt_time}

In principle,
PSR~\psr\ may be directly detectable in these data
through its pulsations.  Because
any emission from the pulsar is embedded in the surrounding extended
component, pulsations will inevitably be contaminated by this extended
emission, and the signal-to-noise ratio of the pulses will thus depend on the
extraction radius and energy range in which data are considered.  We
therefore searched for pulsations by considering events in five
possible circular extraction regions, of radii $2\farcs5$, $5''$,
$10''$, $15''$ and $20''$, each centered on the pulsar position. For each
extraction radius, we considered photons in the energy ranges 0.5--2,
2--5, 5--10 and 0.5--10~keV. We then determined the phase of each
photon by folding the data at the ephemeris listed in Table~\ref{tab_psr}, which
was kindly supplied to us by M. Kramer from pulsar timing observations
at the Jodrell Bank Observatory. Because our two observations were only
separated by two days, we combined data from both runs when searching
for pulsations.

With the phases corresponding to the known
pulsar period in hand, we applied to the data the $Z_n^2$ test
(\cite{bbb+83}), a binning-independent method of looking for the $n$th
harmonic of a pulsating signal in sparsely-sampled data. In the absence
of pulsations, the quantity $Z_n^2$ will be distributed like $\chi^2$
with $2n$ degrees of freedom. For each of the extraction radii and
energy ranges listed above, we have calculated $Z_n^2$ for our
observations for harmonics in the range $1 \le n \le 5$. Amongst these
data, the most significant statistic is $Z_2^2=9.5$, which occurs for
an extraction radius of $15''$ and an energy range 0.5--10~keV. The
probability of such a signal emerging by chance is 4.9\%, which is
consistent with chance occurrence
given the number of different extraction radii and
energy ranges searched. We therefore conclude that there is no
significant pulsed signal in the data.

We can estimate an upper limit on the amplitude of any pulsations as
follows.  For a sinusoidal signal of pulsed fraction $f$ embedded in an
event list containing $N$ photons, we expect $Z_1^2 = 0.5f^2N$
(\cite{lew83}).  Since we do not know the optimal extraction radius in
which pulses are most likely to dominate over surrounding extended
emission, we conservatively consider a $20''$ extraction radius. This
corresponds to $\sim75\%$ of the energy of a central point
source\footnote{See
http://heasarc.gsfc.nasa.gov/docs/xmm/uhb/XMM\_UHB/node17.html .}, but
also most likely suffers from significant contamination from
surrounding emission.  The resulting values of $Z_1^2$ can be used to
determine upper limits on $f$ for the entire core region.  For 0.5--2,
2--5, 5--10 and 0.5--10~keV, we determine upper limits on $f$ for a
sinusoidal pulsed signal from PSR~\psr\ of 11\%, 4\%, 14\% and 2\%
respectively.  The upper limits on the pulsed fraction for an unresolved
pulse are a factor of two lower in each case.

\subsubsection{Imaging Analysis}
\label{pt_image}

The intensity profiles in Figure~\ref{fig_core_slice} show that while
cuts through the pulsar along a north-south axis are consistent with a
point source, those in an east-west direction demonstrate the presence
of extended emission immediately surrounding the pulsar. However,
unlike other cases in which an unresolved source is clearly superposed
on a smooth nebula (e.g.\ \cite{mss+02}), there is no obvious
decomposition of the profile seen 
in the upper panel of Figure~\ref{fig_core_slice} into
two such separate components. The profile shown could be interpreted as
a centrally peaked nebula, with no central point source, but it could
equally be regarded as an unresolved source embedded in lower level
nebular emission. Thus we must adopt a conservative upper limit on a
point source, in which we assume that all the flux in the central bin
corresponds to unresolved emission. Integrating under the appropriately
normalized PSF, we find that in the energy range 0.5--10~keV,
$\sim1010$ counts in the combined MOS1$+$MOS2 data set could be from an
unresolved source at the pulsar position. (Note that only $\sim75\%$ of
these counts would fall within the $18''$ extraction radius shown in
Figure~\ref{fig_mos_zoom}, owing to the large wings of the \xmm\ PSF.)
The corresponding upper limit on the EPIC MOS count-rate from the pulsar is
0.0123~cts~s$^{-1}$.

\subsubsection{Spectral Analysis}
\label{pt_spectrum}

The spectrum of the core region is well fit by a power law. 
If the emission from the pulsar has a blackbody spectrum, then it should
be spectrally distinct from the rest of the core.  Thus, we can
constrain the temperature of a central source by determining what
contribution a thermal source could make to the core's spectrum
(\cite{shm02}). Specifically, we adopt a hard upper limit on the
foreground column density of $N_H < 2 \times 10^{22}$~cm$^{-2}$, above
all values of $N_H$ listed in Table~\ref{tab_spec}.  We assume that the
neutron star's radius as viewed by a distant observer is $R_\infty =
12$~km, and that the distance to the source is 4~kpc.  We can then fit
a two component spectral model to the core's spectrum (as measured by
the combination of MOS1, MOS2 and pn), in which we allow the photon
index and flux of the power-law component to be free parameters, and
find the maximum temperature of the blackbody component which still
yields an acceptable fit to the spectrum. This results in an upper
limit on the surface temperature of the blackbody (as viewed at
infinity) $T_\infty < 147$~eV, and an unabsorbed bolometric luminosity
$L^{bol}_{\infty} < 8.7\times10^{33}$~erg~s$^{-1}$.  We note that a
blackbody model will overestimate a neutron star's temperature when
compared to more realistic atmosphere models (\cite{pz00}),
strengthening the upper limits made from blackbody fits here.

\section{Discussion}
\label{sec_discuss}

\subsection{Comparison with Earlier Results}

{\em ROSAT}\ observations of this source suggested the presence of
three components of emission: an unresolved source centered on the
pulsar, embedded in a compact region of radius $20''$, and further
surrounded by a diffuse nebula of extent $\sim5'$ (\cite{fsp96}). From
the {\em ROSAT}\ data, there was the suggestion that this latter
diffuse component had a cometary morphology, with a fan-like tail
extending out behind the pulsar at a position angle $\sim210^\circ$
(north through east).  In the data presented here, we confirm the
presence of both a compact region surrounding the pulsar, and of a
diffuse nebula at larger extents. However, our results differ from
those of Finley \etal\ (1996\nocite{fsp96}) in two important respects.

First, based on the analyses in \S\ref{pt} above, we can only put upper
limits on the contribution of a central unresolved source.  Given the
many more counts detected here, we ascribe the apparent detection of
such a source by {\em ROSAT}\ to statistical fluctuations in the
presence of low signal-to-noise.  We can estimate the expected
contribution from the pulsar as follows.  The X-ray spectrum of the
Vela pulsar can be modeled as a blackbody with $R_\infty
= 2.1$~km, $T_\infty = 128$~eV and $L^{bol}_{\infty} = 1.5
\times10^{32}$~erg~s$^{-1}$, combined with a power law component of
photon index $\Gamma = 2.7$ and an unabsorbed 0.2--8~keV luminosity
$L_X = 4.2\times10^{31}$~erg~s$^{-1}$ (\cite{pzs+01}).  The
pulsed fraction in the 0.1--10~keV band is 7\% (\cite{hgh01}).
At a distance of 4~kpc, and adopting $N_H = 1\times10^{22}$~cm$^{-2}$
as representative of the minimum of the range of possible absorbing
columns derived in Table~\ref{tab_spec}, we would expect to detect
$\sim35$ EPIC~pn counts in our observation from such a source. When
embedded in the surrounding core region, the effective pulsed fraction
would be $<1\%$, well below the limits determined in \S\ref{pt_time}.
The resulting MOS count-rate for such a source would be
$\sim4\times10^{-4}$~cts~s$^{-1}$, also well below the upper limit on
the point-source count-rate determined in \S\ref{pt_image}. Finally,
the expected temperature and luminosity of the blackbody component are
both much less than the spectral limits on such a source determined in
\S\ref{pt_spectrum}.  Thus the failure to detect PSR~\psr\ in our data
is entirely consistent with the emission properties expected of a
Vela-like pulsar.  Higher resolution observations with {\em
Chandra}\ will be needed to separate out any unresolved source from the
surrounding extended component, and to thus directly detect the pulsar.

Second, while we confirm the presence of diffuse emission surrounding
the core region, we find no evidence for the cometary morphology
proposed by Finley \etal\ (1996\nocite{fsp96}). Rather, it seems
likely this appearance resulted from the much lower sensitivity of
their data ($\sim280$ total source counts for {\em ROSAT}\ PSPC
compared to $\sim4200$ for \xmm\ EPIC MOS), and simply corresponded to
the brightest regions of the larger source seen here.

Our measurements also differ somewhat from the {\em ASCA}\ results
presented by Sakurai \etal\ (2001\nocite{skts01}). While 
Sakurai \etal\ (2001\nocite{skts01}) concluded
that the X-ray emission consists of a core region embedded in
a more diffuse nebula, the diffuse component in their data seems
symmetrically distributed around the pulsar, as opposed to
the distinctly one-sided morphology seen here.  Sakurai
\etal\ (2001\nocite{skts01}) find that the core and diffuse components
can both be fit with power-law spectra, with photon indices of $\Gamma
= 1.9\pm0.2$ and $\Gamma=2.0\pm0.1$ respectively. While these spectra
are broadly consistent with our measurements, the unabsorbed flux
densities (2--10~keV) for the core and diffuse regions inferred from
{\em ASCA}\ are $\sim2\times10^{-11}$~erg~cm$^{-2}$~s$^{-1}$ and
$\sim2\times10^{-12}$~erg~cm$^{-2}$~s$^{-1}$ respectively, an order of
magnitude higher than those inferred here.  Since Sakurai
\etal\ (2001\nocite{skts01}) provide no information on their analysis
procedure, fitted absorbing column or assumed distance, it is difficult
to assess whether there is a genuine discrepancy between our results and
theirs. 
Pending a more detailed analysis of the {\em ASCA}\ data, we
ascribe these differences to the  much poorer angular resolution and
sensitivity of {\em ASCA}\ when compared to \xmm.

\subsection{Morphology of the X-ray Emission}
\label{sec_morph}

Spectral fits to the the diffuse component of the X-ray emission cannot
distinguish between thermal and non-thermal models. In the case of a
thermal spectrum, this emission could be interpreted as shock-heated
gas from a SNR associated with the pulsar.  However, the temperatures
inferred by the corresponding fits listed in Table~\ref{tab_spec} are
higher than seen even in very young ($<2000$~yr) SNRs
(e.g.\ \cite{bh96}; \cite{gv97}).  Fits with enhanced abundances can
lower the inferred temperature (e.g\ \cite{pkc+00}) --- we have
attempted to model the data in this way here, but the quality of the
fit then becomes unacceptable. We thus conclude that the spectrum of
the diffuse component is most likely non-thermal, and is best described
by either the power-law or synchrotron cut-off spectral model.

The synchrotron cut-off spectrum (model ``SRCUT'' in Table~\ref{tab_spec})
corresponds to synchrotron emission from shock-accelerated electrons in
supernova remnants (\cite{rk99}). It is not unreasonable to suppose that a
SNR associated with PSR~\psr\ is accelerating such electrons, and
generating the diffuse component of the emission.  Indeed, the photon
index of a power law fit to the spectrum, $\Gamma \approx 2.3$, is
similar to that seen from other SNRs emitting synchrotron X-rays when
fit with such a model (Slane \etal\ 1999, 2001\nocite{sgd+99,she+01}). 
Furthermore, the inferred 1~GHz
flux density from the SRCUT fit, $15\pm1$~mJy, is well below the upper
limit on 1.4-GHz radio emission from this source ($\sim50$~mJy, $3\sigma$) as
determined by Braun \etal\ (1989\nocite{bgl89}) from their radio
observations of this field. However, in this interpretation the
morphology of the X-ray emission should take the form of an
approximately circular limb-brightened region associated with the
expanding SNR shock (e.g.\ \cite{kpg+95}).  
The amorphous, centrally-brightened emission seen
here does not meet this expectation.

We therefore interpret the diffuse component of the X-rays seen
surrounding PSR~\psr\ as synchrotron emission produced by a pulsar wind
nebula, which we designate G18.0--0.7 based on its Galactic
coordinates.  The fact that the diffuse component steadily fades with
increasing distance from the pulsar supports the interpretation that
PSR~\psr\ is the central energy source for this emission.  The inferred
photon index for the diffuse component, $\Gamma \sim 2.3$, is typical
of PWNe (\cite{bt97}; \cite{pccm02}). The unabsorbed 0.5--10~keV
luminosity for this source is $L_X \sim
3d_4^2\times10^{33}$~erg~s$^{-1}$; the corresponding efficiency with
which the pulsar's spin-down is converted into nebular X-rays is
$L_X/\dot{E} = 1.1\times10^{-3}$, a value also typical for young
pulsars (\cite{bt97}).

Contained completely within the diffuse component, the core region is a
compact region with a hard power-law spectrum centered on the pulsar.
Bright central structures are seen around many other young pulsars
which power PWNe (e.g.\ \cite{hgh01}; \cite{gak+02}), and similarly we
argue here that the core region is a central region of synchrotron
emission powered by the pulsar.  We thus interpret the core and diffuse
regions of X-ray emission as both parts of a single PWN, of total
unabsorbed luminosity (0.5--10~keV) $L_X \approx
4d_4^2\times10^{33}$~erg~s$^{-1}$ and foreground column density $N_H
\approx 1.2\times10^{22}$~cm$^{-2}$.  Spectral fits with the ``SRCUT''
model demonstrate that a continuous synchrotron spectrum for G18.0--0.7
extrapolated to radio frequencies results in a 1-GHz flux density well
below the sensitivity of radio searches for a nebula of this size,
accounting for the non-detection of any PWN surrounding PSR~\psr\ at
radio wavelengths (\cite{bgl89}; \cite{gsf+00}).

In such a source, we expect that the pulsar wind flows freely out to
the point at which the wind pressure is balanced by that from the
nebula.  In this region the wind decelerates, a termination shock is
formed, and synchrotron emission is generated. Recent high-resolution
observations with the {\em Chandra X-ray Observatory}\ have provided
beautiful images of termination shocks around young pulsars. In some
cases these images demonstrate that the termination region is not a
spherical shock, but rather has a toroidal morphology presumably
corresponding to a wind focused into the pulsar's equatorial plane
(\cite{gak+02}; \cite{lwa+02}).

While we generally expect any X-ray PWN to be brightest near the
pulsar, Figure~\ref{fig_core_slice} demonstrates that the core is a
distinct, well-defined region superimposed on the much fainter diffuse
background.  We thus argue that the core component seen here represents
the pulsar termination shock, its elongated morphology possibly
indicative of anisotropic outflow from the pulsar. The radius of the
termination shock is then $r_s \approx 15'' = 0.3d_4$~pc, an extent
similar to that seen around the Crab pulsar and PSR~B1509--58
(\cite{wht+00}; \cite{gak+02}), but about an order of magnitude larger
than that inferred for the Vela pulsar (\cite{hgh01}).  We expect
particles emitting in this region to have undergone minimal synchrotron
losses, and to thus have a harder spectrum than the rest of the nebula
(e.g.\ \cite{scs+00}; \cite{gak+02}).  We can directly compare the
spectra of the core and diffuse components by fitting each to a power
law with fixed $N_H$. The results of such fits are given in
Table~\ref{tab_spec} for $N_H = 1.0\times10^{22}$~cm$^{-2}$
(corresponding to the best-fit absorbing column for the core) and $N_H
= 1.4\times10^{22}$~cm$^{-2}$ (corresponding to the best fit for the
diffuse region).  From these fits, we see that the core is indeed
harder than the rest of the nebula at a statistically significant
level, with a difference in photon index $\Delta\Gamma \approx
0.3-0.5$.

The X-ray luminosity of the
core region is about 25\% of that from the total nebula. This is
somewhat higher than the $\sim10\%$ contribution made by the central
termination shock region in other such sources (\cite{scs+00};
\cite{gak+02}; \cite{lwa+02}), potentially indicating a difference in
the efficiency or length scale over
which electrons and positrons are accelerated in the shock seen
here compared to other sources. 
However, we caution that a significant contribution to the flux
in the core region could be from the pulsar itself (see \S\ref{pt_image}
above).

If the core indeed represents the termination shock,
the extent of this region then provides a constraint on the magnetic field
in G18.0--0.7. Specifically, we expect pressure balance
between the pulsar wind and the surrounding nebula,
\begin{equation}
P_{neb} \approx \frac{\dot{E}}{4\pi\Omega r_s^2 c}, 
\end{equation}
where $P_{neb}$ is the pressure in the PWN
and $\Omega$ is the fraction of a sphere
through which the wind flows.  The limited angular resolution of
\xmm\ makes it difficult to derive a specific geometry
for outflow from the core morphology; we here leave $\Omega$
as a free parameter with a hard upper limit $\Omega \le 1$. The corresponding
nebular pressure needed to confine the pulsar wind is $P_{neb} =
9.2\Omega^{-1}d_4^{-2}\times10^{-12}$~erg~cm$^{-3}$.
At distances far downstream of
the shock, we expect approximate equipartition between particles and
magnetic fields (\cite{kc84a}), so that $B^2/8\pi \approx 0.5P_{neb}$,
where $B$ is the mean nebular magnetic field. In this case we can infer
$B \approx 10\Omega^{-1/2}d_4^{-1}$~$\mu$G.  For electron/positron pairs emitting at 1.5~keV (the
peak of the X-ray flux seen in Figure~\ref{fig_pwn_spec}), the corresponding
synchrotron lifetime is $t_{synch} \sim1000\Omega^{3/4}d_4^{3/2}$~yr.

The simplest expectation for a PWN is that, beyond the termination
shock, the emitting particles should flow out radially in all
directions until they either radiate all their energy or reach the
outer edge of the PWN. However, G18.0--0.7 differs from this in two
important respects. Most fundamentally, rather than forming an
approximate ellipsoid with the pulsar and termination shock at its
center (e.g.\ Slane \etal\ 2000, 2002\nocite{scs+00,shm02}), G18.0--0.7
is distinctly one-sided, showing little emission to the north of the
pulsar but a broad diffuse region to the south.  Also problematic is
the fact the southern region of the PWN extends to a distance
$L\sim5'=6d_4$~pc from the pulsar.  If this extent represents the outer
edge of the expanding nebula in this direction, then the expansion
speed of the nebula is $V \approx L/\tau \approx
300d_4$~\kms\ (e.g.\ \cite{vagt01}).  Alternatively, if the true
nebular extent is much larger than that seen in X-rays, then we expect
a flow speed at the edge of the X-ray nebula $V \approx (r_s/L)^2(c/3)
\approx 250$~\kms\ (\cite{kc84a}). Both these calculations are at odds
with the speed needed to transport particles to the southern end of the
PWN within their synchrotron lifetimes, $L/t_{synch} \approx
6000\Omega^{-3/4}d_4^{-1/2}$~\kms. Thus the size of the PWN is much
larger than be easily accounted for if the nebula has expanded
unimpeded on its southern side.

One possible explanation for the PWN's properties is if if the pulsar's
space velocity is supersonic with respect to the sound speed of its
environment. In this case the PWN is confined through ram pressure and
a bow-shock PWN results (e.g.\ \cite{gjs02}). In such PWNe, the pulsar
is at one end of an elongated X-ray nebula; the PWN has a bright head
at or near the pulsar position, with a progressively fainter tail
extending opposite the direction of motion (e.g.\ \cite{kggl01}).
While there is no proper motion information for PSR~\psr\ from which we
might estimate its space velocity, the gross morphology of G18.0--0.7
could be interpreted as a tail extending to one side of the pulsar,
suggestive of rapid motion of the central source in a northerly
direction.  However, there are at least two difficulties
with this interpretation.

First, in other bow-shock PWNe, the shocked pulsar wind interior to the
bow shock is tightly confined by the flow of surrounding material, and
the X-ray tail correspondingly consists of a very narrow column of
emission, marginally resolved or unresolved even with {\em
Chandra}\ (\cite{buc02}; \cite{sgk+02}).  On the other hand, the
diffuse emission seen here forms a very broad region behind the pulsar,
showing no evidence for containment or collimation.
Second, the length of the proposed tail region seen here, $L \approx
6d_4$~pc, is too long to be a trail of emission left behind by a fast
pulsar.  Even if the pulsar had an extreme space velocity $V_{PSR} = 2000$~\kms,
the lifetime computed above for X-ray emitting pairs limits the length
of such a synchrotron ``wake'' to $V_{PSR}~t_{synch}
\sim2\Omega^{3/4}d_4^{3/2}$~pc,
noticeably smaller than the measured extent of G18.0--0.7.
Alternatively, if the emitting particles have not been left behind by
the pulsar, but are part of the wind flow driven by the pulsar, one is
left with the uncomfortably high-flow speed discussed above,
$L/t_{synch} \approx 6000$~\kms.  Wang \etal\ (1993\nocite{wlb93})
argue that such high-speed backflows behind the pulsar can be attained
when the wind is forced to travel through a narrow nozzle, but no such
collimated region behind the pulsar is seen here.

A second possibility to explain the morphology and extent of this PWN
is that the nebula is intrinsically symmetric around the pulsar, but
that there is a relativistic bulk flow of particles within the nebula
which results in Doppler boosting of the approaching component over
receding material.  Such a situation has been proposed to explain the
one-sided PWN seen around PSR~B1509--58 (\cite{gak+02}), in which a
flow velocity $\beta = v/c \ga 0.2$ has been inferred. However, in PWNe
for which such high speed flows have been claimed, this outflow
manifests itself in collimated jets of material directed along the
pulsar spin-axis (\cite{gak+02}; \cite{lwa+02}) --- the lack of any
such collimated structures here argues against this possibility.  The
large extent of the PWN can be explained in this case, since the flow
time from the pulsar to the edge of the nebula is only $t_{flow}
\approx 20\beta^{-1}$~yrs (ignoring projection effects).  However, at
the relativistic speeds $\beta \ga 0.1$ needed for Doppler boosting to
be effective,  we find that $t_{flow} \ll t_{synch}$. We thus would
expect minimal synchrotron cooling of the outer nebula (i.e.\ a photon
index $\Gamma \approx 1.6$ as is seen in the core), in contradiction
with the steeper photon index ($\Gamma \approx 2.3$) seen for the
diffuse component of G18.0--0.7.

A final alternative is that the system is in an evolved state, in which
the reverse shock from a surrounding SNR has propagated back inwards,
colliding with the PWN (\cite{rc84}).  If different components of this
reverse shock collide with the PWN at different times, emitting
particles can be rapidly convected away from the pulsar, generating a
nebula which can be significantly offset to one side of the pulsar
(\cite{che98}).  Such an effect has been specifically proposed to
account for the one-sided morphology of the PWN powered by the Vela
pulsar (\cite{bcf01}), and we propose that this process may similarly
be occurring here for the Vela-like pulsar~\psr. Such an asymmetric PWN
will result only if the PWN crushing time-scale is of order or less
than the lifetime of emitting pairs, and if the velocity of convective
motions is sufficient to produce the observed PWN extent.  Simulations
of this process suggest a crushing time-scale $\sim2000$~yrs and
convective motions at speeds $\sim1000$~\kms (\cite{vagt01}).  Assuming
that the PWN had a radius of a few pc before interacting with the
reverse shock, these values are roughly consistent with the observed
nebular extent and the lifetimes of emitting electrons inferred above,
and we thus think an asymmetric reverse-shock interaction the most
likely explanation for the observed nebular morphology.  One important
consideration in this interpretation is that we do not directly detect
the SNR required to generate this effect; we discuss this issue in
\S\ref{sec_no_snr} below.

Recent evidence has suggested that characteristic ages can overestimate
(\cite{krv+01}) or underestimate (\cite{gf00}) a pulsar's true age by
up to an order of magnitude, depending on assumptions made about the
initial period of and braking torque acting on the neutron star.  The
PWN / reverse shock interaction which we have proposed here is expected
to take place $\sim10^4$~yrs after the supernova explosion
(\cite{rc84}), in approximate agreement with the characteristic age
$\tau \equiv P/2\dot{P} = 21.4$~kyr measured for PSR~\psr. Thus within
the reverse shock scenario, the available evidence suggests that the
pulsar's characteristic age and true age are broadly consistent.

\subsection{The Lack of an Associated Supernova Remnant}
\label{sec_no_snr}

The interpretation which we have adopted in \S\ref{sec_morph} requires
the presence of a surrounding SNR in order to generate the reverse
shock which consequently interacts with the PWN. However, no such SNR
has been detected at radio wavelengths (\cite{bgl89}), nor do we see
any surrounding shell of X-ray emission from such a source in the data
presented here.

The lack of detectable SNRs around pulsars with small characteristic
ages has long been problematic (e.g.\ \cite{ra85}; \cite{fkcg95}). It
is thought that such SNRs must exist, as otherwise the PWN would expand
freely into the ambient medium, and would produce a highly extended and
unobservable ``ghost nebula'' (\cite{bopr73}).  In the specific case of
Vela-like pulsars such as PSR~\psr, Braun \etal\ (1989) interpret the
absence of a detectable radio SNR as resulting from the rapid expansion
of the SNR blast wave into a low density cavity, followed by a
collision with a dense surrounding shell evacuated by the progenitor
wind.  The SNR's radio and X-ray emissions then fade rapidly as the
SNR's energy is dissipated through radiative shocks. Simulations of
this process clearly show that the SNR/shell interaction drives a
reverse shock back towards the center of the system
(e.g.\ \cite{tbfr90}). Furthermore, either an offset of the progenitor
from the center of its wind bubble or pressure variations within the
turbulent SNR/bubble interaction region can cause different parts of
the reverse shock to collide with the PWN at different times
(\cite{rtfb93}; \cite{dwa01}), compressing the PWN in a highly
asymmetric fashion as is required here.

We can consider whether such a SNR would be detected by scaling the
emission seen from the Vela SNR. Lu \& Aschenbach (2000\nocite{la00})
find that X-rays from the Vela SNR can be modeled as a two-component
Raymond-Smith plasma, with temperatures for the two components of $kT_1
\approx 0.15$~keV and $kT_2 \approx 1$~keV.  The volume-integrated
emission measure for the cooler component is $\sim10$ times that of the
hotter component, and the total unabsorbed luminosity (0.1--2.5~keV) is
$2.2\times10^{35}$~erg~s$^{-1}$ for a distance of 250~pc. Scaling to a
distance of 4~kpc, the expected unabsorbed 0.1--2.5~keV flux densities
for the two components are $f_1 =
1.0\times10^{-10}$~erg~cm$^{-2}$~s$^{-1}$ and $f_2 =
9.5\times10^{-12}$~erg~cm$^{-2}$~s$^{-1}$.  

Using the W3PIMMS
tool,\footnote{http://asc.harvard.edu/toolkit/pimms.jsp} we can
determine that the expected EPIC MOS count rate from such a source is
0.18--0.29~cts~s$^{-1}$ for absorbing columns in the range
$(1.0-1.4)\times10^{22}$~cm$^{-2}$. Again scaling from the Vela SNR, we
expect this emission to be distributed around a shell of diameter
$\sim20'$. Assuming a shell thickness 20\% of its radius, the expected
surface brightness for EPIC MOS would be
$(4-7)\times10^{-7}$~cts~s$^{-1}$~arcsec$^{-2}$, several times below
the observed background level $10'$ off-axis of
$\sim3\times10^{-6}$~cts~s$^{-1}$~arcsec$^{-2}$ (after correction for
vignetting). We can make similar calculations for archival observations
of this source with {\em ROSAT}\ and {\em ASCA}: for {\em ROSAT} PSPC
we expect a SNR surface brightness
$(5-9)\times10^{-8}$~cts~s$^{-1}$~arcsec$^{-2}$ against an observed
background in these data of
$\sim2\times10^{-7}$~cts~s$^{-1}$~arcsec$^{-2}$, while for {\em
ASCA}\ GIS we expect a SNR surface brightness
$(4-6)\times10^{-8}$~cts~s$^{-1}$~arcsec$^{-2}$ against the observed
background of $\sim2\times10^{-7}$~cts~s$^{-1}$~arcsec$^{-2}$.  Thus we
expect that any emission from a Vela-like SNR associated with
PSR~\psr\ would not be detectable in any of these observations.  This
conclusion is strengthened by the fact that the characteristic age of
PSR~\psr\ is approximately twice that of the Vela pulsar; if this
indicates a similar difference in these pulsars true ages, any
associated SNR would most likely be even fainter than considered here.

For the radio observations carried out in this direction by Braun
\etal\ (1989\nocite{bgl89}), we estimate a 1-GHz surface brightness
sensitivity for a SNR of this extent of $\Sigma \ga
1.0\times10^{-21}$~W~m$^{-2}$~Hz$^{-1}$~sr$^{-1}$ (3 $\sigma$), where
we have assumed a typical radio photon index $\Gamma = 1.5$ and have
accounted for primary beam attenuation. There are certainly many SNRs
fainter than this, including relatively young SNRs such as G266.2--1.2
(\cite{dg00}).  Thus, the failure to detect a SNR associated with
PSR~\psr\ is not inconsistent with the possibility that a rapidly
expanding, rapidly fading SNR has produced the pressure which has
confined and distorted the PWN.

\section{Conclusions}

We have carried out a deep X-ray observation of the Vela-like pulsar
\psr\ with \xmm. We find that all the detected X-ray emission results
from the interaction between the relativistic pulsar wind and its
environment: a bright elongated core immediately surrounding the pulsar
most likely represents the wind termination shock, while a surrounding
faint region with a softer spectrum corresponds to a pulsar wind nebula
with a mean
magnetic field of $\sim10$~$\mu$G. Emission from the pulsar itself,
either in the form of a central point source or as pulsations at the
101-ms pulsar period, is not detected.

The nebula, which we designate G18.0--0.7, is distinctly one-sided: most of its emission 
is distributed in a region of extent $5'$ restricted to the southern side
of the pulsar. We interpret this morphology as resulting from the
compression and distortion of the pulsar wind by the asymmetric reverse
shock produced by an unseen surrounding supernova remnant.  Deeper
observations of PSR~\psr\ and other young pulsars at meter wavelengths,
carried out with the Very Large Array and with LOFAR, are probably the
most promising way to directly detect these associated SNRs.  The
reverse shock interaction between a pulsar nebula and its surrounding
supernova remnant is expected to occur 10--20~kyr after the
supernova explosion, suggesting that many Vela-like pulsars might be in
this phase of evolution. Future observations of other such
pulsars and their nebulae with \xmm\ can explore this possibility.

\begin{acknowledgements}

We thank Herman Marshall for assistance with the observing proposal,
Michael Kramer for providing a timing ephemeris for PSR~\psr, and
Patrick Slane for helpful discussions.  The results presented are based
on observations obtained with \xmm, an ESA science mission with
instruments and contributions directly funded by ESA Member States and
the USA.  B.M.G. acknowledges the support of NASA through \xmm\ Guest
Observer grant NAG5-11376, and of a Clay Fellowship awarded by the
Harvard-Smithsonian Center for Astrophysics. V.M.K. is an Alfred P.\
Sloan Fellow and a Canada Research Chair.

\end{acknowledgements}


\clearpage

\begin{table}[hbt]
\caption{\xmm\ observations of PSR~\psr.}
\label{tab_obs}
\begin{tabular}{lcccccc} \hline
Date of Observation & Length of Observation (ks)  & \multicolumn{3}{c}{Effective Exposure (ks)} \\
 & & MOS1 & MOS2 & pn \\ \hline
2001 October 16 & 30.5 & 17.8 & 18.3 & 10.2  \\
2001 October 18 & 29.8 & 23.0 & 23.2 & 12.6  \\
TOTAL & 60.3 & 40.8 & 41.5 & 22.8 \\ \hline
\end{tabular}

\tablenotetext{}{Note that the effective exposures have been corrected for dead-time effects.}
\end{table}

\begin{table}[hbt]
\caption{Radio timing ephemeris for PSR~\psr.}
\label{tab_psr}
\begin{tabular}{lcccccc} \hline
RA (J2000)  & $18^{\rm h}26^{\rm m}13\fs175(2)$ \\
Dec. (J2000) & $-13^\circ 34' 46\farcs 7(2)$ \\
Frequency (Hz) & 9.85564535524(1) \\
Frequency derivative (Hz s$^{-1}$) & $-7.2930848(2)\times10^{-12}$ \\
Frequency second derivative (Hz s$^{-2}$) & $1.51530(6)\times10^{-22}$ \\
Epoch of Ephemeris (MJD) & 50750.000000 \\
Range of Validity (MJD) & 49156 to 52359 \\ \hline
\end{tabular}

\tablenotetext{}{These data have been supplied by M. Kramer from timing observations with Jodrell Bank Observatory. Numbers
in parentheses indicate the uncertainty in the last significant
figure of each value.}
\end{table}

\begin{table}[hbt]
\caption{Spectral fits to X-ray emission from G18.0--0.7.}
\label{tab_spec}
\scriptsize
\begin{tabular}{lccccccccc} \hline
Region & \multicolumn{3}{c}{Total Counts} & Model & $N_H$ & $\Gamma/kT/E_0$ & $f_X$ & $\chi_\nu^2/\nu$ \\
       & MOS1 & MOS2 & pn & & ($10^{22}$~cm$^{-2}$) &  ---/(keV)/(keV)
 & ($10^{-13}$~erg~cm$^{-2}$~s$^{-1}$)\tablenotemark{a} \\ \hline
Core\tablenotemark{b} & $455\pm23$ & $458\pm22$ & $815\pm33$ & PL & $1.0\pm0.2$ &
$1.6^{+0.1}_{-0.2}$ & $3.5^{+1.0}_{-0.8}$  & $42/38 = 1.1$ \\
 & & & & PL & 1.0 (fixed) & $1.6\pm0.1$ & $2.9\pm0.3$  & $42/39 = 1.1$ \\
 & & & & PL & 1.4 (fixed) & $1.8\pm0.1$ & $3.0\pm0.3$  & $49/39 = 1.3$ \\
   & & & & BB & $<0.2$ & $1.19\pm0.08$ & $2.1\pm0.5$  & $60/38=1.6$ \\ 
\\ \hline
Diffuse  & $1843\pm135$ & $2360\pm138$ & $\ldots$ & PL & $1.4^{+0.5}_{-0.2}$ 
& $2.3^{+0.4}_{-0.3}$ & $18_{-7}^{+12}$  & $130/133=1.0$ \\
Component & & & & PL & 1.0 (fixed) & $1.9\pm0.1$ & $13\pm2$  & $135/134 = 1.0$ \\
 & & & & PL & 1.4 (fixed) & $2.3^{+0.1}_{-0.2}$ & $17\pm3$  & $130/134 = 1.0$ \\
 & & &  & Brem & $1.1^{+0.4}_{-0.2}$ & $3.9^{+2.2}_{-1.1}$ &
$12^{+4}_{-3}$ & $129/133=1.0$ \\ 
& & & & SRCUT & $1.4\pm0.2$ & $\sim1$ & $15\pm1$ & $130/133=1.0$ \\
 & & &  & RS & $1.3^{+0.2}_{-0.3}$ & $3.4^{+1.3}_{-0.8}$ &
$13^{+3}_{-2}$  & $133/133=1.0$ \\
& & & & MEKAL & $1.3^{+0.2}_{-0.3}$ & $3.4^{+1.2}_{-0.8}$ &
$13^{+3}_{-2}$  & $136/133=1.0$ \\
& & &  & NEI & $1.2^{+0.3}_{-0.2}$ & $3.8^{+1.3}_{-1.1}$ &
$38^{+5}_{-9}$  & $126/132=1.0$ \\
\hline
\end{tabular}
\normalsize

\tablenotetext{}{Uncertainties are all at 90\% confidence.
All models assume interstellar absorption using the
cross-sections of Ba\protect\mbox{\l}uci\protect\'{n}ska-Church 
\& McCammon (1992\protect\nocite{bm92}), assuming solar abundances.
Models used: ``PL'' indicates a power law of the
form $f_\varepsilon \propto \varepsilon^{-\Gamma}$ where $\Gamma$ is the
photon index; ``BB'' indicates a blackbody emitting
at a temperature $T$; ``Brem'' indicates thermal
bremsstrahlung; ``SRCUT'' indicates a synchrotron spectrum from a power-law
distribution of electrons in a homogeneous magnetic field, with
an exponential cut-off at a characteristic energy $E_0$
(\protect\cite{rk99}); 
``RS'' indicates a Raymond-Smith spectrum of
temperature $T$ (\protect\cite{rs77}); ``MEKAL'' indicates
a hot plasma model at temperature $T$;
``NEI'' indicates a non-equilibrium ionization plasma at
temperature $T$ with solar abundances.}
\tablenotetext{a}{Flux densities are for the energy range 0.5--10~keV,
and have been corrected for interstellar absorption.}
\tablenotetext{b}{Counts and flux densities listed here for the core 
correspond only to $\sim75\%$ of the total for
this region (see text for details).} 
\end{table}

\end{document}